\documentclass[aps,pre,amsmath,lengthcheck,superscriptaddress]{revtex4-2}

\usepackage{graphicx}\graphicspath{ {figures/} }
\usepackage{hyperref}
\hypersetup{colorlinks,allcolors=blue,breaklinks}

\newcommand{\be}{\begin{equation}}
\newcommand{\ee}{\end{equation}}
\newcommand{\bs}{\begin{split}}
\newcommand{\es}{\end{split}}
\newcommand{\ba}{\begin{align}}
\newcommand{\ea}{\end{align}}
\newcommand{\bi}{\begin{itemize}}
\newcommand{\ei}{\end{itemize}}

\newcommand{\la}{\left\langle}
\newcommand{\ra}{\right\rangle}
\newcommand{\pd}{\partial}

\newcommand{\bla}{bla\\bla\\bla\\bla\\bla}

\newcommand{\mb}[1]{\mbox{\boldmath$#1$}}
\newcommand{\mc}[1]{\mathcal{#1}}

\begin{document}

\title{Hamilton's equations for relaxation function}

\author{Pierre Naz\'e}
\email{pierre.naze@unesp.br}

\affiliation{\it Universidade Federal do Par\'a, Faculdade de F\'isica, ICEN,
Av. Augusto Correa, 1, Guam\'a, 66075-110, Bel\'em, Par\'a, Brazil}

\date{\today}

\begin{abstract}

The relaxation function is the cornerstone to perform calculations in weakly driven processes. Properties that such a function should obey are already established, but the difficulty in its calculation is still an issue to be overcome. In this work, I proposed a new method to determine such a function for thermally isolated systems, based on a Hamilton's equations approach. Observing that the microscopic relaxation function can be turned into a canonical variable, one can choose the initial conditions of the solutions of Hamilton's equations to avoid the calculation of the average in the initial canonical ensemble. The unbearable example of the quartic oscillator is solved to corroborate the method. Extensions to the quantum realm and stochastic thermodynamics are mandatory.

\end{abstract}

\maketitle

\section{Preliminaries}
\label{sec:preliminaries}

I start by defining notations and developing the main concepts to be used in this work.

Consider a classical system with a Hamiltonian $\mc{H}(\mb{z}(\mb{z_0},t)),\lambda(t))$, where $\mb{z}(\mb{z_0},t)$ is a point in the phase space $\Gamma$ evolved from the initial point $\mb{z_0}$ until time $t$, with $\lambda(t)$ being a time-dependent external parameter. Initially, the system is in thermal equilibrium with a heat bath of temperature $\beta\equiv {(k_B T)}^{-1}$, where $k_B$ is Boltzmann's constant. After that, the system is decoupled from the bath, and during a switching time $\tau$, the external parameter is changed from $\lambda_0$ to $\lambda_0+\delta\lambda$. The average work performed on the system during this interval of time is
\be
\langle W\rangle = \int_0^\tau \la\pd_{\lambda}\mc{H}(t)\ra_0\dot{\lambda}(t)dt,
\label{eq:work}
\ee
where $\partial_\lambda$ is the partial derivative in respect to $\lambda$ and the superscripted dot the total time derivative. The generalized force $\la\pd_{\lambda}\mc{H}\ra_0$ is calculated using the averaging $\langle\cdot\rangle_0$ over the initial canonical ensemble. The external parameter can be expressed as
\be
\lambda(t) = \lambda_0+g(t)\delta\lambda,
\label{eq:ExternalParameter}
\ee
where, to satisfy the initial conditions of the external parameter, the protocol $g(t)$ must satisfy the following boundary conditions
\be
g(0)=0,\quad g(\tau)=1. 
\label{eq:bc}
\ee
I consider as well that $g(t)\equiv g(t/\tau)$, which means that the intervals of time are measured according to the switching time unit.

Linear-response theory aims to express average quantities until the first-order of some perturbation parameter considering how this perturbation affects the observable to be averaged and the process of average \cite{kubo2012statistical}. In our case, I consider that the parameter does not considerably change during the process, $|g(t)\delta\lambda/\lambda_0|\ll 1$, for all $t\in[0,\tau]$. In that manner, using such a framework, the average work produced is~\cite{naze2020compatibility}
\begin{equation}
\begin{split}
\langle W\rangle = &\, \delta\lambda\la\pd_{\lambda}\mc{H}\ra_0-\frac{\delta\lambda^2}{2}(\Psi_0(0)-\la\pd_{\lambda\lambda}^2\mc{H}\ra_0)\\
&+\frac{\delta\lambda^2}{2} \int_0^\tau\int_0^\tau \Psi_0(t-t')\dot{g}(t')\dot{g}(t)dt'dt.
\label{eq:work}
\end{split}
\end{equation}
where $\Psi_0(t)$ is the relaxation function, given by
\be
\Psi_0(t) = \beta\la\pd_\lambda\mc{H}(0)\pd_\lambda\mc{H}(t)\ra_0-\mc{C},
\ee 
where $C$ is a constant. Such a function is the cornerstone to perform the calculations in weakly driven processes. Other features, such as optimal protocols, depend almost exclusively on the relaxation function. Knowing how to calculate $\Psi_0(t)$ is therefore mandatory.

\section{Constant $\mc{C}$}

About the constant $C$, two options are possible: if
\be
\mc{C}=\beta\langle\partial_\lambda H\rangle_0^2,
\ee
then 
\be
(\Delta F)_2 = \delta\lambda\la\pd_{\lambda}\mc{H}\ra_0-\frac{\delta\lambda^2}{2}(\Psi_0(0)-\la\pd_{\lambda\lambda}^2\mc{H}\ra_0),
\ee
where $\Delta F$ is the difference of Helmholtz's free energy between the final and initial equilibrium states of the system. However, if
\be
\mc{C}=\beta\langle\partial_\lambda H\rangle_0^2+2(\langle W_{\rm qs}\rangle-\Delta F)_2,
\ee
then
\be
(\langle W_{\rm qs}\rangle)_2 = \delta\lambda\la\pd_{\lambda}\mc{H}\ra_0-\frac{\delta\lambda^2}{2}(\Psi_0(0)-\la\pd_{\lambda\lambda}^2\mc{H}\ra_0),
\ee
where $\langle W_{\rm qs}\rangle$ is the quasistatic work of the system.

\section{Usual method}

The calculation of the relaxation function has some steps:

\begin{itemize}
    \item Calculate the solution of the non-perturbed Hamilton's equations;
    \item Take an average on the initial canonical ensemble.
\end{itemize}
If Hamilton's equations have an explicit analytical solution, the problem is manageable, although it becomes difficult to take the average if the initial conditions are not linearly dependent on the solutions. In the case when it is not, the problem becomes hard to calculate. Indeed, initial conditions must be sampled according to the initial canonical ensemble, the equations numerically solved with them, and the relaxation function becomes a numerical average over them at the end. Observing that such a process holds only for one $\tau$, then a whole analysis of the non-equilibrium dynamics becomes almost impossible to obtain. 

\section{New method}

The new method proposed has the advantage of not performing the average anymore. 
To start, consider the following type of Hamiltonian
\be
\mc{H}(q(t),p(t),\lambda(t))=\frac{p(t)^2}{2}+V(q(t),\lambda(t))
\ee
Under weakly driven processes, the effective Hamiltonian is the non-perturbed one, given by
\be
\mc{H}_0(q_0,p_0,\lambda_0)=\frac{p_0^2}{2}+V(q_0,\lambda_0).
\ee
Calling the microscopic relaxation function $\psi_0$ as a new canonical variable
\be
\psi_0 = \beta\partial_{\lambda_0}V(a,\lambda_0)\partial_{\lambda_0}V(q_0,\lambda_0),
\label{eq:psi0}
\ee
one needs to find its canonical conjugated $\theta_0$. Such a function can be calculated by using the condition
\be
\{\psi_0,\theta_0\}=1,
\ee
where $\{\cdot,\cdot\}$ is the Poisson bracket on the canonical variables $(q_0,p_0)$. Solving the equation, one has
\be
\theta_0 = \frac{1}{\partial_{q_0}\psi_0}p_0+f(q_0),
\label{eq:theta0}
\ee
where $f(q_0)$ is an arbitrary function. For simplicity, I consider $f(q_0)=0$. Solving Eqs.~\eqref{eq:psi0} and~\eqref{eq:theta0} to $q_0$ and $p_0$, one has
\be
q_0=G_1(\psi_0,\theta_0),\quad p_0=G_2(\psi_0,\theta_0).
\ee
Substituting in the non-perturbed Hamiltonian, one has
\be
\mc{H}_0(\psi_0,\theta_0)=\frac{G_2(\psi_0,\theta_0)^2}{2}+\partial_{\lambda_0}V(G_1(\psi_0,\theta_0),\lambda_0),
\ee
whose Hamilton's equations
\be
\dot{\psi_0}=\partial_{\theta_0}\mc{H}_0,\quad \dot{\theta_0}=-\partial_{\psi_0}\mc{H}_0,
\ee
furnishes the solutions $\psi_0(a,t)$ and $\theta_0(a,t)$, which depend on $\psi_0(0)$ and $\theta_0(0)$. In particular, the number $a$ is chosen to satisfy
\be
\psi_0(0)=\beta\partial_{\lambda_0}V(a,\lambda_0)^2.
\ee
The relaxation function is expected to depend only on $\langle\psi_0(0)\rangle_0$. In this manner, I use the following argument: taking the average on the initial canonical ensemble in $\psi_0(t)$ is the same thing as choosing $\psi_0(0)$ and $\theta_0(0)$ such that the following restriction is achieved
\be
\langle \psi_0(t)\rangle_0=\langle \psi_0(0)\rangle_0 (C(t)-\mc{C}).
\ee
In general, this is achieved by choosing 
\be
\psi_0(0)=\langle \psi_0(0)\rangle_0, \quad \theta_0(0)=0.
\ee
The relaxation function finally is
\be
\Psi_0(t) = \langle \psi_0(0)\rangle_0 (C(t)-\mc{C}).
\ee
Remark that by using such a method, the part of solving differential equations remains, but taking the average in the initial canonical ensemble is avoided. Also, the function $\langle\psi_0(0)\rangle_0$ has no relevance to the general characteristic of the irreversible work and optimal control procedures since it can be taken as a natural unit to work with. The constant $\mc{C}$, according to the examples provided, is the one associated with $\Delta F$. This result is not yet understood.

\subsection{Harmonic oscillator}

Let us recover the relaxation function of the harmonic oscillator perturbed in the stiffness parameter $\lambda(t)$. Consider its non-perturbed Hamiltonian
\be
\mc{H}_0(q_0,p_0,\lambda_0)=\frac{p_0^2}{2}+\lambda_0\frac{q_0^2}{2} 
\ee
The microscopic relaxation function $\psi_0$ is
\be
\psi_0 = \frac{\beta}{4}a^2q_0^2 
\ee
The canonical conjugate is
\be
\theta_0 = \frac{2p_0}{\beta a^2 q_0}
\ee
Solving for $q_0$ and $p_0$, one has
\be
q_0 = \frac{2}{\sqrt{\beta} a}\sqrt{\psi_0}, \quad p_0 = \sqrt{\beta} a \theta_0\sqrt{\psi_0}.
\ee
Substituting in the non-perturbed Hamiltonian, one has
\be
\mc{H}_0(\psi_0,\theta_0) = \left(\frac{\beta a^2 \theta_0^2}{2}+\frac{2\lambda_0}{\beta a^2}\right)\psi_0.
\ee
The Hamilton's equations are
\be
\dot{\psi_0}=\beta a^2\theta_0\psi_0, \quad \dot{\theta_0} = -\left(\frac{\beta a^2 \theta_0^2}{2}+\frac{2\lambda_0}{\beta a^2}\right).
\ee
Using
\be
a= \left(\frac{4\psi_0(0)}{\beta}\right)^{1/4},\quad  \psi(0)=\langle \psi_0(0)\rangle_0,\quad \theta_0(0)=0, 
\ee
the solution is
\be
C(t) -\mc{C}= \cos(\sqrt{\lambda_0}t)^2.
\ee
Adapting $\mc{C}$ to the quasistatic work, the relaxation function will be
\be
\Psi_0(t)=\Psi_0(0) \cos{(2\sqrt{\lambda_0}t)}.
\ee
which agrees with the relaxation function calculated in Ref.~\cite{acconcia2015degenerate}.

\subsection{Quartic oscillator}

\begin{figure}
    \centering
    \includegraphics[scale=0.8]{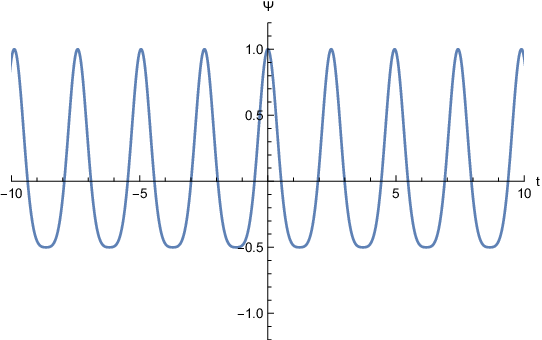}
    \caption{Normalized relaxation function for a quartic oscillator. It was used $\beta=1$ and $\lambda_0=1$.}
    \label{fig1}
\end{figure}

\begin{figure}
    \centering
    \includegraphics[scale=0.8]{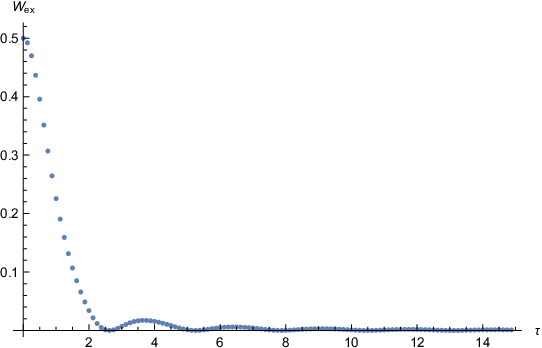}
    \caption{Excess work for a quartic oscillator. It was used $\beta=1$, $\lambda_0=1$ and $\delta\lambda=0.1$.}
    \label{fig2}
\end{figure}

Consider the non-perturbed Hamiltonian of the quartic oscillator driven in the stiffness parameter
\be
\mc{H}_0(q_0,p_0,\lambda_0)=\frac{p_0^2}{2}+\lambda_0\frac{q_0^4}{4} .
\ee
The microscopic relaxation function $\psi_0$ is
\be
\psi_0 = \frac{\beta}{16}a^4q_0^4.
\ee
The canonical conjugate is
\be
\theta_0 = \frac{4p_0}{\beta a^4 q_0^3}.
\ee
Solving for $q_0$ and $p_0$, one has
\be
q_0 = \frac{2}{\beta^{1/4} a}\psi_0^{1/4}, \quad p_0 = 2\beta^{1/4} a \theta_0\psi_0^{3/4}.
\ee
Substituting in the non-perturbed Hamiltonian
\be
\mc{H}_0(\psi_0,\theta_0) = 2\beta^{1/2} a^2 \theta_0^2\psi_0^{3/2}+\frac{4\lambda_0}{\beta a^4} \psi_0.
\ee
Hamilton's equations are
\be
\dot{\psi_0}=4\beta^{1/2} a^2 \theta_0\psi_0^{3/2}, \quad \dot{\theta_0} = -\left(3\beta^{1/2} a^2 \theta_0^2\psi_0^{1/2}+\frac{4\lambda_0}{\beta a^4}\right).
\ee
Using,
\be
a= \left(\frac{16\psi_0(0)}{\beta}\right)^{1/8},\quad \psi_0(0)=\langle\psi_0(0)\rangle_0,\quad \theta_0=0,
\ee
and calculating $\mc{C}$ for the quasistatic work, the relaxation function can be found numerically. Observe that the relaxation function is a periodic function by the phase space configuration of the system. The normalized relaxation function is depicted in Fig.~\ref{fig1}. It is an even function, which can be expanded into a cosine series with coefficients verified to be non-negative until the 10th order. This indicates that its Fourier transform is probably non-negative. Also, numerical calculation indicates that the waiting time, given by
\be
\tau_w = \mc{L}_s\left[\frac{\Psi_0(t)}{\Psi_0(0)}\right](0),
\ee
where $\mc{L}_s$ is the Laplace transform, is null for the same reason. This is expected to thermally isolated systems~\cite{naze2023shortcuts}. Also, this indicates that the system has shortcuts to adiabaticity for all $\tau$~\cite{naze2023shortcuts}. In Fig.~\ref{fig2}, the excess work is numerically calculated for the linear protocol. The strong inequality $\langle W\rangle\ge \langle W_{\rm qs}\rangle$ holds for it. The excess work decreases to zero for large $\tau$, as expected.

\bibliography{HERF}

\end{document}